\documentclass[12pt]{article}
\usepackage{a4wide,amssymb,cite}
\usepackage{a4wide,amssymb,cite,graphicx}
\usepackage{epsfig}
\usepackage[usenames,dvipsnames]{color}
\usepackage{slashed}
\pdfoutput=1

\usepackage{a4wide,amssymb,cite,graphicx}
\usepackage[usenames,dvipsnames]{color}
\usepackage{slashed}
\usepackage{amsfonts}
\usepackage{amsmath}
\usepackage[small]{caption}
\usepackage[multiple]{footmisc}

\renewcommand{\tilde}{\widetilde}
\newcommand{\be}{\begin{equation}}
\newcommand{\ee}{\end{equation}}
\newcommand{\bea}{\begin{eqnarray}}
\newcommand{\eea}{\end{eqnarray}}

\def\circa#1{\,\raise.3ex\hbox{$#1$\kern-.75em\lower1ex\hbox{$\sim$}}\,}

\allowdisplaybreaks

\begin{document}

\begin{titlepage}
\def\thefootnote{\fnsymbol{footnote}}

\begin{flushright}
  KIAS-P15005\\
  APCTP Pre2015-002\\
  IPMU15-0009
\end{flushright}

\begin{centering}
\vspace{1cm}
{\Large {\bf Supersymmetric Higgs-portal and X-ray lines}} \\
%
\bigskip

\begin{center}
  {\bf
    Hyun~Min~Lee$^{a,}$\footnote{Email: hminlee@cau.ac.kr},
    Chan~Beom~Park$^{b,}$\footnote{Email: cbpark@kias.re.kr} and
    Myeonghun~Park$^{c,d,e,}$\footnote{Email: parc@apctp.org}}
\end{center}

\begin{center}
  {\em $^a$Department of Physics, Chung-Ang University, Seoul
    156-756, Korea}\\[0.2cm]
  {\em $^b$School of Physics, Korea Institute for Advanced Study,
    Seoul 130-722, Korea}\\[0.2cm]
  {\em $^c$Asia Pacific Center for Theoretical Physics,\\
    77 Cheongam-Ro, Nam-Gu, Pohang 790-784, Korea}\\[0.2cm]
  {\em $^d$Department of Physics, Postech, Pohang 790-784, Korea}\\[0.2cm]
  {\em $^e$Kavli IPMU (WPI), The University of Tokyo, Kashiwa, Chiba
    277-8583, Japan}\\[0.2cm]
\end{center}

\bigskip

\end{centering}
%

\begin{abstract}
\noindent
We consider a Dirac singlet fermion as thermal dark matter for explaining the X-ray line in the context of a supersymmetric Higgs-portal model or a generalized Dirac NMSSM. The Dirac singlet fermion gets a mass splitting due to their Yukawa couplings to two Higgs doublets and their superpartners, Higgsinos, after electroweak symmetry breaking. We show that a correct relic density can be obtained from thermal freeze-out, due to the co-annihilation with Higgsinos for the same Yukawa couplings. We discuss the phenomenology of the Higgsinos in this model such as displaced vertices at the LHC.

\end{abstract}

\vspace{3cm}

\end{titlepage}

\renewcommand{\thefootnote}{\arabic{footnote}}
\setcounter{footnote}{0}

\section{Introduction}
\noindent
Dark matter (DM) is a main component of matter in the Universe, confirmed by various observations such as galaxy rotation curves, gravitational lensing. Moreover, it is supported by the
measurement of Cosmic Microwave Background Radiation and Large Scale Structure, and so on. However, we have no clue as to the DM mass and interactions other than gravity. Therefore, direct detection on earth, indirect detection in the sky, and direct production at particle colliders have been thought to be complementary for identifying the nature of DM. In particular, indirect detections look for the remnants of annihilations or decays of DM through cosmic rays coming from galaxies and galaxy clusters.

There has recently been a lot of interest in light DM models, after new detection of X-ray line coming from galaxies and galaxy clusters mainly by the XMM-Newton observatory~\cite{xray}. There are on-going debates on the possibility of explaining the X-ray line excess with thermal atomic transition~\cite{debate}, but it is worthwhile to take it to be a signal for DM and study the consequences of decaying or annihilating DM models~\cite{dmmodels,mdm,mdmothers,falkowski}.

Motivated by a toy model suggested by one of us~\cite{mdm}, we consider a concrete model for explaining the X-ray line with the magnetic dipole moment of a weakly interacting massive particle (WIMP) in the context of a generalized next-to-minimal supersymmetric standard model (NMSSM) with an additional Dirac singlet superfield, dubbed as Dirac NMSSM~\cite{dnmssm,kai}. Unlike the toy model where a discrete $Z_2$ symmetry for stabilizing DM is broken by a small amount
at the cutoff scale~\cite{mdm}, the corresponding discrete parity, i.e. $R$-parity, in the supersymmetric (SUSY) version is assumed to be exact. Then, a singlet Dirac fermion or two Majorana fermions called the singlinos, introduced in the Dirac NMSSM, is the DM candidate, and it gets a small mass splitting for the X-ray line energy at $3.55\,{\rm keV}$ due to its small Yukawa couplings to the MSSM Higgses and their superpartners. In this case, a tiny magnetic transition dipole moment for decaying DM generates the X-ray line by the small Yukawa couplings of the singlinos. We regard the model as a SUSY Higgs-portal in the limit that gauginos, squarks and sleptons are heavy enough. We also include the effects of non-decoupled gauginos on the mass splitting of Higgsinos or singlinos. The lightness of Higgsinos and singlinos can be ensured by a chiral symmetry such as Peccei-Quinn symmetry while gauginos could be relatively light due to $R$-symmetry.  

The Dirac singlet fermion can keep in thermal equilibrium with the Standard Model (SM) particles at freeze-out, due to the co-annihilation with the Higgsino-like fermions. Consequently, we show that the correct relic density can be attained, being compatible with the X-ray line. In the limit of heavy gauginos, the mass splitting of Higgsino states is about ${\rm keV}$ scale as for the singlino fermions, so neutral or charged Higgsinos decay into a singlino $+Z^\ast / W^\ast$, leaving a displaced vertex due to small Yukawa couplings of singlinos. We discuss the possibility of discovering Higgsinos at the LHC in this new topology.

The paper is organized as follows.
We begin with the model description of the SUSY Higgs-portal for the low-energy mass spectra of DM and Higgsinos. Then, we present the results of the magnetic transition dipole moment between two singlinos at one loop in our model and show the parameter space that is consistent with both the energy and flux for the X-ray line. In turn, we discuss the bound from the DM relic density and its compatibility with the X-ray line. Finally, conclusions are drawn.

\section{Supersymmetric Higgs-portal}
\noindent
The dark sector couples to the SM particles only through the Higgs and its superpartners. As an example, we consider an extension of the Higgs sector in the MSSM with a Dirac singlet chiral superfield containing two additional singlet superfields, $S$ and $\bar S$. We assume that the gauginos as well as the superpartners of quarks and leptons are sufficiently heavy so that they are not relevant for our discussion. Meanwhile, we also discuss the effects of non-decoupled gauginos in this section.

The part of the superpotential containing only Higgs doublets, $H_u$ and $H_d$, and the singlet chiral superfields are
\be
W_0=  \lambda_S S H_u H_d +  \lambda_{\bar S} {\bar S} H_u H_d+ M_S S {\bar S} + \mu_H H_u H_d +  \mu_S S + \mu_{\bar S} {\bar S}.
\ee
In this model, the Dirac singlet chiral superfield communicates with the SM only through the Higgs and Higgsino interactions. As for the Dirac singlino, the model can also be called the Higgino portal.
In a Peccei-Quinn (PQ) symmetric realization of the above superpotential, the cubic couplings for the singlet chiral superfields are forbidden, while the bare Higgsino and singlino mass terms and the singlet tadpole terms can be generated after a spontaneous breaking of the PQ symmetry by non-renormalizable interactions with PQ-breaking fields.

When there is a $U(1)_{\rm S}$ global symmetry or a $Z_2$ symmetry distinguishing $S$ and ${\bar S}$, the operator ${\bar S} H_u H_d$ is forbidden. This case corresponds to the Dirac NMSSM that was discussed in Ref.~\cite{dnmssm,kai}, where even after integrating out the singlet scalar masses with keeping their fermion partners, the resulting Higgs potential gets an additional quartic potential, $|\lambda_S H_u H_d|^2$, and increases the Higgs mass as compared to the MSSM.
When the singlet symmetry is broken spontaneously or explicitly, we can write a small Yukawa coupling for $\bar S$ such that $|\lambda_{\bar S} |\ll |\lambda_S|={\cal O}(1)$. Then, the feature of the Dirac NMSSM for the Higgs mass can be maintained.

On the other hand, if $|\lambda_S|$ and $|\lambda_{\bar S}|$ are comparable, the PQ symmetry only does not distinguish between $S$ and ${\bar S}$. Thus, there is no obvious reason to forbid Majorana mass terms such as $S^2$ and ${\bar S}^2$ in the superpotential.
But, if we ignore those Majorana mass terms under the assumption that such a flavor structure in the dark sector is determined by a flavor symmetry for singlinos at a high energy scale, we can explain a small mass splitting and a small flux required for the X-ray line for $|\lambda_S|$, $|\lambda_{\bar S}|\ll 1$, as will be discussed in the next section.

The neutralino mass matrix containing the gauginos in MSSM is given in the basis $({\tilde B},\, {\tilde W}^0,\, {\tilde H}^0_d,\, {\tilde H}^0_u,{\tilde S},\, {\tilde {\bar S}})$ by
\be
M_{{\tilde\chi}^0}= \left(\begin{array}{cccccc} M_1 & 0 &  -\frac{1}{2}g^\prime v_d & \frac{1}{2} g^\prime v_u  & 0 & 0\\
0 & M_2 & \frac{1}{2}g v_d  & -\frac{1}{2}g v_u & 0 & 0 \\
-\frac{1}{2}g^\prime v_d & \frac{1}{2}g v_d & 0 & -\mu_{\rm eff} & -\frac{1}{\sqrt{2}} \lambda_S v_u & -\frac{1}{\sqrt{2}} \lambda_{\bar S} v_u \\
\frac{1}{2}g^\prime v_u & -\frac{1}{2}g v_u & -\mu_{\rm eff} & 0 & -\frac{1}{\sqrt{2}} \lambda_S v_d & -\frac{1}{\sqrt{2}} \lambda_{\bar S} v_d \\
0 & 0 & -\frac{1}{\sqrt{2}} \lambda_S v_u & -\frac{1}{\sqrt{2}} \lambda_S v_d & 0 & M_S \\
0 & 0 & -\frac{1}{\sqrt{2}} \lambda_{\bar S} v_u & -\frac{1}{\sqrt{2}} \lambda_{\bar S} v_d & M_S & 0
  \end{array} \right) ,
\ee
where $v_u^2+v^2_d=v^2 \simeq (246\,{\rm GeV})^2$, $\tan\beta=v_u/v_d$, and the effective $\mu$ parameter is given by $\mu_{\rm eff}=\mu_H+ \lambda_S \langle S\rangle+\lambda_{\bar S} \langle {\bar S}\rangle$.

In order to keep a small mass splitting between singlinos, we take the gauginos to be much heavier than Higgsinos and singlinos, namely, $M_{1,2}\gg \mu_{\rm eff}, M_S$. Then, we can consider only the $4\times 4$ sub-matrix for Higgsinos and singlinos
and a mass splitting of the Dirac singlinos is attributed to a small coupling between Higgsinos and singlinos. Then, keeping all the other superpartners of the SM heavy enough, we can call the model the SUSY Higgs-portal.

\begin{figure}
  \begin{center}
    \includegraphics[height=0.48\textwidth]{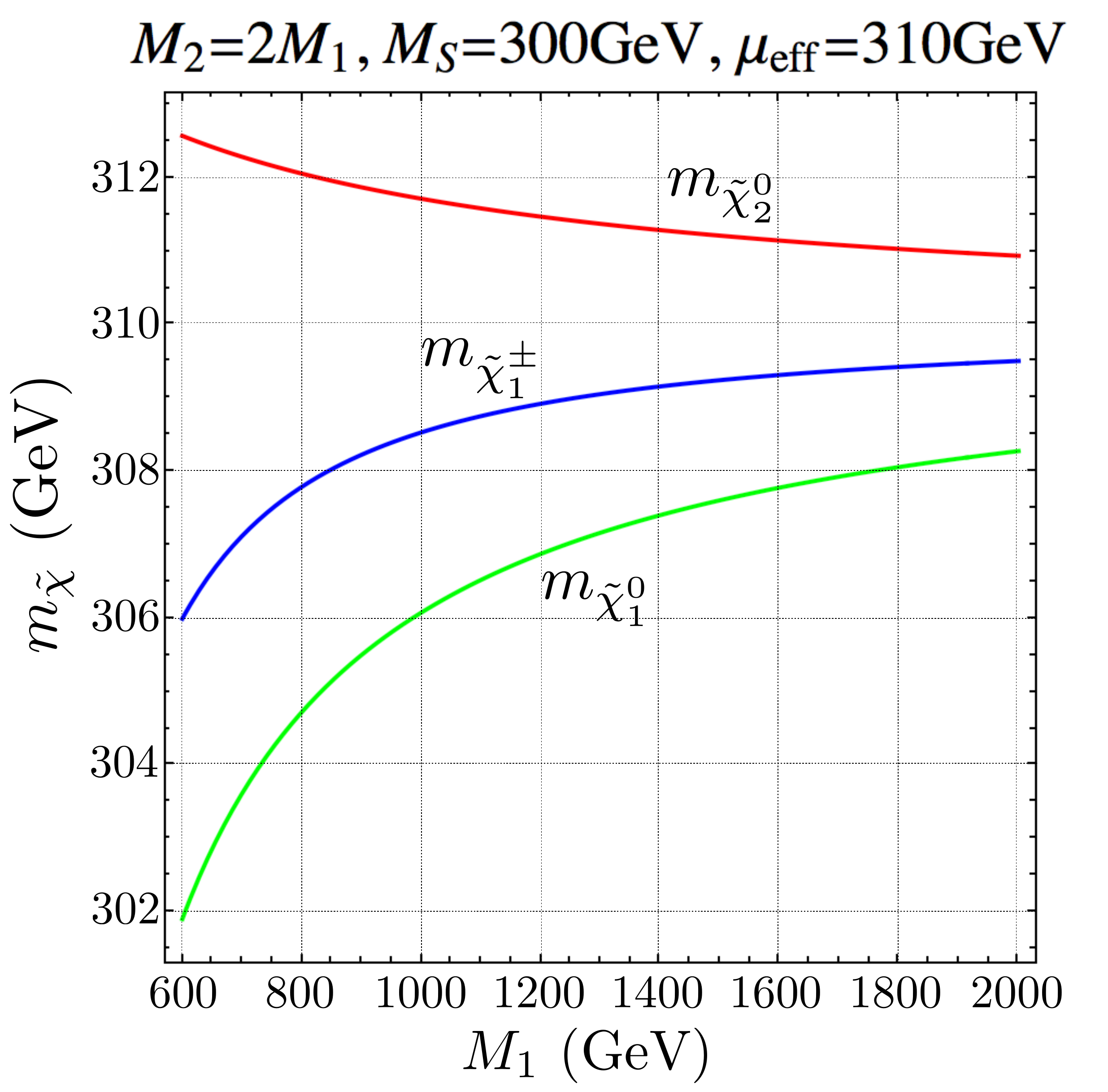}
  \end{center}
  \caption{ Masses of Higgsino-like states as a function of gaugino mass $M_1$.
    We ignored the mixing effect with singlinos and took $\tan\beta=10$ and $\mu_{\rm eff}>0$. }
  \label{Higgsinomass}
\end{figure}

In the limit of $M_{1,2}\gg \mu_{\rm eff}, M_S$, the mass eigenvalues for Higgsino-like neutralinos are
\begin{align}
m_{{\tilde\chi^0}_1} &= \mu_{\rm eff}-\frac{1}{8}(v_u+v_d)^2\left(\frac{g^{\prime 2}(M_1-2\mu_{\rm eff})}{(M_1-\mu_{\rm eff})^2}+\frac{g^2(M_2-2\mu_{\rm eff})}{(M_2-\mu_{\rm eff})^2} \right), \nonumber\\
m_{{\tilde \chi}^0_2} &= \mu_{\rm eff}+\frac{1}{8}(v_u-v_d)^2 \left(\frac{g^{\prime 2}(M_1+2\mu_{\rm eff})}{(M_1+\mu_{\rm eff})^2}+\frac{g^2(M_2+2\mu_{\rm eff})}{(M_2+\mu_{\rm eff})^2} \right),
\end{align}
while those for singlino-like neutralinos are, for $\lambda_S,\lambda_{\bar S}\ll 1$,
\begin{align}
m_{{\tilde\chi}^0_3}=&~  M_S+\frac{1}{8}(\lambda_S-\lambda_{\bar S})^2\left(\frac{(v_u-v_d)^2}{\mu_{\rm eff}+M_S}-\frac{(v_u+v_d)^2}{\mu_{\rm eff}-M_S}\right) \nonumber \\
&+\frac{1}{16}(\lambda_S -\lambda_{\bar S})^2\,\frac{(v^2_u-v^2_d)^2 \mu^2_{\rm eff}}{(\mu^2_{\rm eff}-M^2_S)^2}\left(\frac{g^{\prime 2}(M_1+2M_S)}{(M_1+M_S)^2}+\frac{g^2(M_2+2M_S)}{(M_2+M_S)^2}\right) ,\nonumber\\
m_{{\tilde \chi}^0_4} =&~ M_S+ \frac{1}{8}(\lambda_S+\lambda_{\bar S})^2\left(\frac{(v_u+v_d)^2}{\mu_{\rm eff}+M_S}-\frac{(v_u-v_d)^2}{\mu_{\rm eff}-M_S}\right) \nonumber \\
&-\frac{1}{16}(\lambda_S +\lambda_{\bar S})^2\,\frac{(v^2_u-v^2_d)^2 \mu^2_{\rm eff}}{(\mu^2_{\rm eff}-M^2_S)^2}\left(\frac{g^{\prime 2}(M_1-2M_S)}{(M_1-M_S)^2}+\frac{g^2(M_2-2M_S)}{(M_2-M_S)^2}\right).
\end{align}
Consequently, the mass differences between the nearest neutralinos are
\begin{align}
\Delta m_{21} &\equiv m_{{\tilde\chi}^0_2}-m_{{\tilde\chi}^0_1} \nonumber \\
 &\approx \frac{1}{4}v^2\left(\frac{g^{\prime 2}}{M_1}+\frac{g^2}{M_2}\right),
\end{align}
and
\begin{align}
\Delta m_{34} \equiv&~ m_{{\tilde\chi}^0_3}-m_{{\tilde\chi}^0_4} \nonumber  \\
\approx&~ \frac{1}{2}\frac{v^2}{\mu^2_{\rm eff}-M^2_S}\, \Big((\lambda^2_+-\lambda^2_-)M_S-(\lambda^2_++\lambda^2_-)\mu_{\rm eff} \sin(2\beta) \Big) \nonumber \\
&+\frac{1}{8}(\lambda^2_++\lambda^2_-)
  \,\frac{v^4\cos^2(2\beta)\mu^2_{\rm eff}}{(\mu^2_{\rm
  eff}-M^2_S)^2}\,\left(\frac{g^{\prime
  2}}{M_1}+\frac{g^2}{M_2}\right) ,
\end{align}
where
\be
\lambda_\pm\equiv  \frac{1}{\sqrt{2}}(\lambda_S \pm \lambda_{\bar S}).
\ee

\begin{figure}
\begin{center}
  \includegraphics[height=0.48\textwidth]{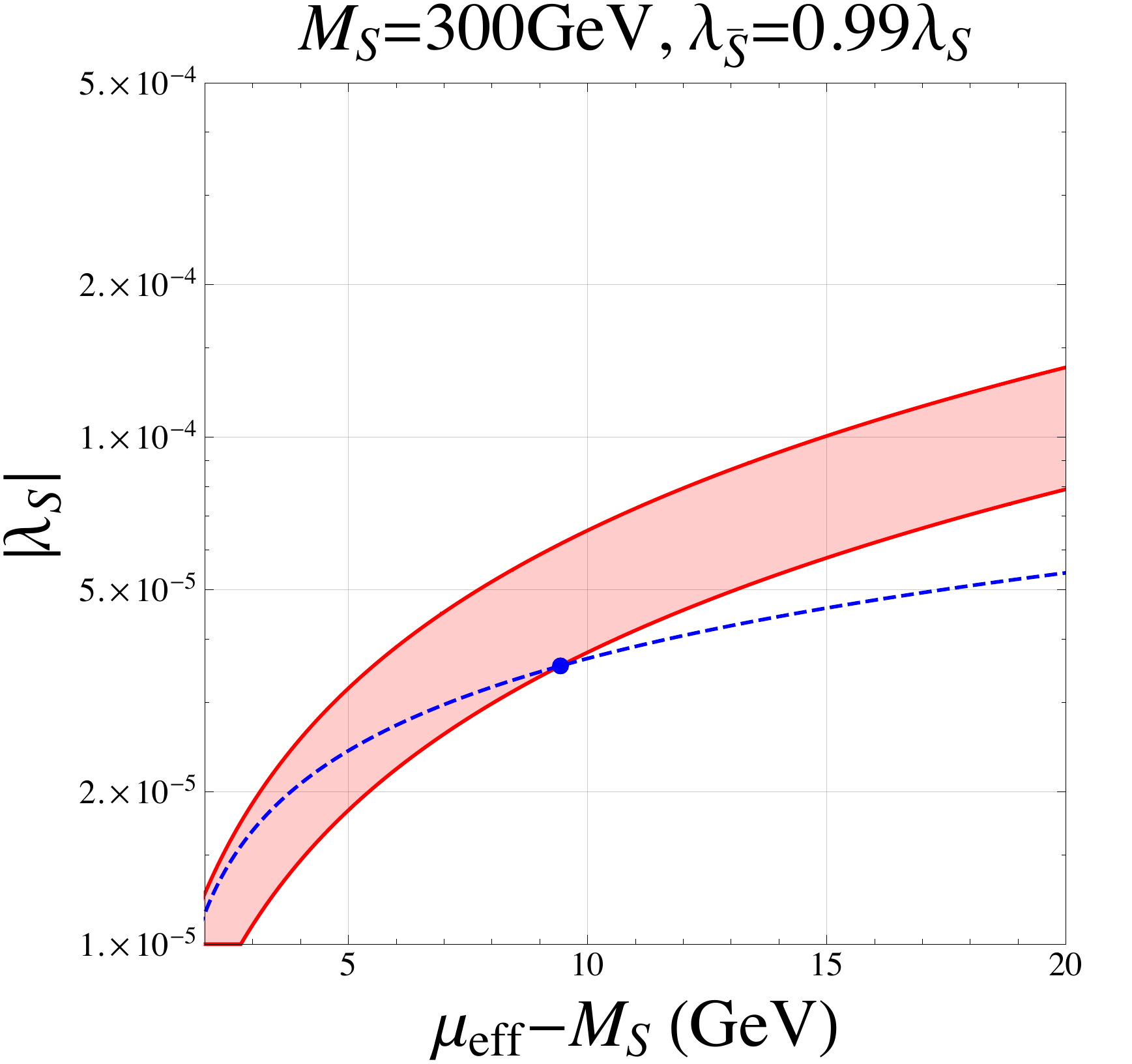}
  \includegraphics[height=0.48\textwidth]{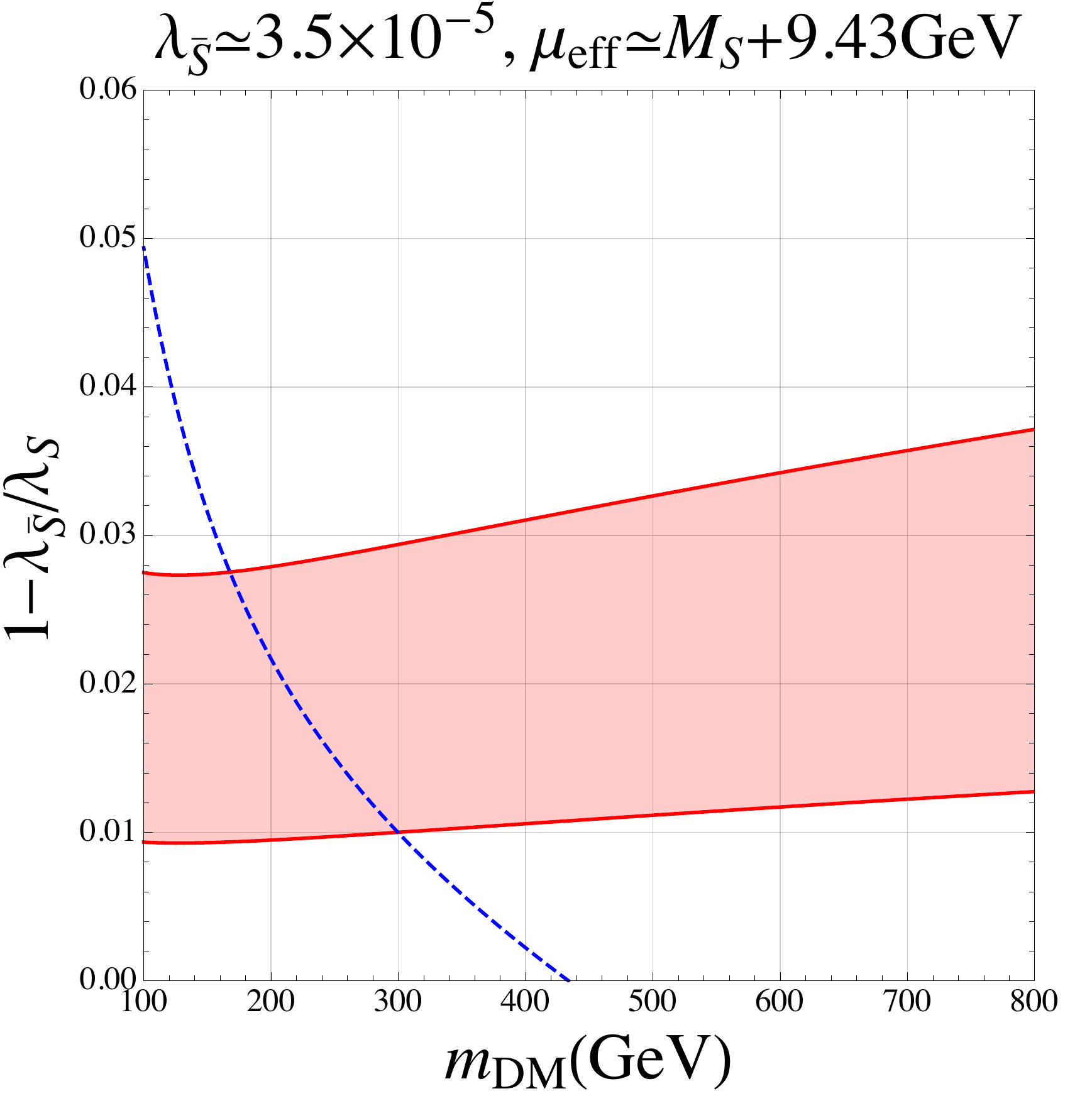}
\end{center}
\caption{(Left) Parameter space of $\mu_{\rm eff}-M_S$ vs $|\lambda_S|$. (Right) Parameter space of $\mu_{\rm eff}-M_S$ vs $1-\lambda_{\bar S}/\lambda_S$. In both figures, the parameter space explaining the X-ray line is shown between two red solid lines and the X-ray line energy at $3.55\,{\rm keV}$ is obtained for the blue dashed line. We took $m_{H^\pm}=1\,{\rm TeV}$, $M_1=0.5 M_2=3\,{\rm TeV}$ and $\tan\beta=10$.}
\label{allbounds}
\end{figure}

We note that as far as $\lambda_S$ and $\lambda_{\bar S}$ are comparable, $\Delta m_{34}$ is positive so that ${\tilde\chi}^0_4$ is the Lightest Supersymmetric Particle (LSP) and ${\tilde\chi}^0_3$ is Next-LSP in our model.  
When the singlino mass splitting is about a few $\,{\rm keV}$ and $\mu_{\rm eff}\gtrsim M_S \sim 100\,{\rm GeV}$, the Yukawa couplings, $\lambda_S$ and $\lambda_{\bar S}$, should be of order $10^{-5}$ and the gaugino masses should be greater than about $1\,{\rm TeV}$, unless there is an accidental cancellation.\footnote{In the case of cancellation, the Yukawa couplings, $\lambda_S$ and $\lambda_{\bar S}$, can be of order one so they can be used to increase the Higgs mass \cite{work2}.}
In Fig.~\ref{Higgsinomass}, we have illustrated the masses of Higgsino-like neutralinos as a function of the gaugino mass. For gaugino masses being greater than 1~TeV and  DM mass being $300\,{\rm GeV}$, the Higgsino mass splitting $\Delta m_{21}$ is less than $6\,{\rm GeV}$. In Fig.~\ref{allbounds}, we show the parameter space for the Yukawa couplings and the mass parameters satisfying the mass splitting between singlino-like neutralinos, $|\Delta m_{34}|=3.55\,{\rm keV}$, in blue dashed line.

The mass eigenstates are found by ${\tilde H}^0_d=\sum_i N_{i1} {\tilde\chi}^0_i$, ${\tilde H}^0_u=\sum_i N_{i2} {\tilde\chi}^0_i$, ${\tilde S}=\sum_i N_{i3} {\tilde\chi}^0_i$ and ${\tilde {\bar S}}=\sum_i N_{i4} {\tilde\chi}^0_i$. For $M_{1,2}\gg \mu_{\rm eff}, M_S$, and $\lambda_S,\,\lambda_{\bar S}\ll 1$, they read
\begin{align}
{\tilde H}^0_d= &~ \frac{1}{\sqrt{2}}  {\tilde\chi}^0_1 + \frac{1}{\sqrt{2}}\,i\gamma^5 {\tilde\chi}^0_2- \frac{\sqrt{2}}{4}\lambda_-\left(\frac{v_u-v_d}{\mu_{\rm eff}+M_S} -\frac{v_u+v_d}{\mu_{\rm eff}-M_S}  \right)i\gamma^5 {\tilde\chi}^0_3 \nonumber \\
&+ \frac{\sqrt{2}}{4} \lambda_+ \left(\frac{v_u-v_d}{\mu_{\rm eff}-M_S} -\frac{v_u+v_d}{\mu_{\rm eff}+M_S}  \right) {\tilde\chi}^0_4,  \label{tildeHd} \\
{\tilde H}^0_u=&~ -\frac{1}{\sqrt{2}}  {\tilde\chi}^0_1 -\frac{1}{\sqrt{2}} \,i\gamma^5 {\tilde\chi}^0_2+\frac{\sqrt{2}}{4}\lambda_- \left(\frac{v_u-v_d}{\mu_{\rm eff}+M_S} +\frac{v_u+v_d}{\mu_{\rm eff}-M_S}  \right)i\gamma^5 {\tilde\chi}^0_3 \nonumber \\
&-\frac{\sqrt{2}}{4}\lambda_+ \left(\frac{v_u-v_d}{\mu_{\rm eff}-M_S} +\frac{v_u+v_d}{\mu_{\rm eff}+M_S}  \right)  {\tilde\chi}^0_4 ,  \label{tildeHu} \\
{\tilde S}=&~-\frac{\sqrt{2}}{4}(v_u-v_d) \left(\frac{\lambda_-}{\mu_{\rm eff}+M_S} +\frac{\lambda_+}{\mu_{\rm eff}-M_S}  \right) {\tilde\chi}^0_1\nonumber \\
&-\frac{\sqrt{2}}{4}(v_u+v_d) \left(\frac{\lambda_+}{\mu_{\rm eff}+M_S} +\frac{\lambda_-}{\mu_{\rm eff}-M_S}  \right)i\gamma^5  {\tilde\chi}^0_2 -\frac{1}{\sqrt{2}}i\gamma^5   {\tilde\chi}^0_3  +\frac{1}{\sqrt{2}} {\tilde\chi}^0_4,  \label{tildeS} \\
{\tilde {\bar S}} =&~ \frac{\sqrt{2}}{4}(v_u-v_d)\left(\frac{\lambda_-}{\mu_{\rm eff}+ M_S}-\frac{\lambda_+}{\mu_{\rm eff}-M_S} \right) {\tilde\chi}^0_1  \nonumber \\
&-\frac{\sqrt{2}}{4}(v_u+v_d)\left(\frac{\lambda_+}{\mu_{\rm eff}+ M_S}-\frac{\lambda_-}{\mu_{\rm eff}-M_S} \right) i\gamma^5 {\tilde\chi}^0_2+ \frac{1}{\sqrt{2}}i \gamma^5 {\tilde\chi}^0_3 +\frac{1}{\sqrt{2}} {\tilde\chi}^0_4.  \label{tildeSbar}
\end{align}

The chargino mass matrix in the basis $({\tilde W}^+, \, {\tilde H}_u^+ , \,
{\tilde W}^- , \, {\tilde H}_d^-)$ is
\be
M_{{\tilde\chi}^\pm}= \left(\begin{array}{cc}    M_2 & \frac{1}{\sqrt{2}}g v_u  \\
    \frac{1}{\sqrt{2}} g v_d & \mu_{\rm eff} \end{array} \right).
\ee
Then, for $M_2\gg \mu_{\rm eff}$, the mass eigenvalues for charginos are
\begin{align}
m_{{\tilde\chi}^\pm_1}&= \mu_{\rm eff} - \frac{\mu_{\rm eff}+M_2 \sin(2\beta)}{M^2_2-\mu^2_{\rm eff}}\,\cdot {\rm sgn}(\mu_{\rm eff}) m^2_W, \nonumber\\
m_{{\tilde\chi}^\pm_2}&= M_2 +\frac{M_2+\mu_{\rm eff} \sin(2\beta)}{M^2_2-\mu^2_{\rm eff}} \, \cdot m^2_W.
\end{align}
The mass difference between the lighter Higgsino-like neutralino  and the lighter chargino is
\be
m_{{\tilde\chi}^\pm_1}-m_{{\tilde\chi}^0_1}= \frac{m^2_W}{2g^2} \left( \Big(1+\sin(2\beta)\Big) \frac{g^{\prime 2}}{M_1}+\Big(1+(1-2{\rm sgn}(\mu_{\rm eff}))\sin(2\beta)\Big) \frac{g^2}{M_2}\right).
\ee
In Fig.~\ref{Higgsinomass}, we have also shown the masses of Higgsino-like chargino as a function of the gaugino mass. In this example, the mass difference between the lighter Higgsino-like neutralino and the Higgsino-like chargino is less than $2.5\,{\rm GeV}$ for gauginos being heavier than $1\,{\rm TeV}$.

Before closing the section, we remark on the scalar sector of the SUSY Higgs-portal.
Due to the small Yukawa couplings of the singlinos, their superpartners, singlet scalars, have only a small mixing with the MSSM Higgs fields so the Higgs sector is MSSM-like. Moreover, it would be hard to produce singlet scalars at the current LHC at a detectable level.  On the other hand, singlet scalars may induce the self-annihilation and co-annihilation of DM through the $s$-channels. The self-annihilation is suppressed due to small Yukawa couplings outside the resonance, while the co-annihilation is sizable enough to keep DM in thermal equilibrium, as will be discussed in Sec.~\ref{sec:relic}.

\section{Magnetic dipole moments and the X-ray line}
\noindent
The magnetic (transition) dipole moments can be obtained for either Majorana~\cite{Nmdm,mdm} or Dirac~\cite{jekim} singlet DM.\footnote{Similar studies on magnetic dipole moments have been done in light of the X-ray line in Ref.~\cite{mdmothers,falkowski}.} In the SUSY Higgs-portal model, the heavier singlino ${\tilde \chi}^0_3$ is a Majorana fermion that has an almost degenerate mass with the lighter singlino ${\tilde \chi}^0_4$. As gauginos, leptons, and squarks are assumed to be decoupled in our model, only charged Higgs and $W$-boson loops contribute to the magnetic dipole moment.
In order to compute the magnetic transition dipole moment for singlinos, we choose the non-linear $R_\xi$ gauge~\cite{haber}, in which $\gamma-W^\pm-G^\mp$ interactions are absent for the charged unphysical Goldstone boson $G^\pm$. In this case, we have to deal with only the Yukawa interactions for Goldstone bosons, thus simplifying the calculation.

The singlino Yukawa interactions with charged Higgs ($H^\pm$) and
charged Goldstone ($G^\pm$) are
\begin{align}
-{\cal L}_S=&~ \sin\beta\, {\overline{ {\tilde\chi}^-_2} }P_L ( \lambda_S{\tilde S}+\lambda_{\bar S}{\tilde {\bar S}} ) H^- + \cos\beta\, {\overline {{\tilde\chi}^+_2}}P_L ( \lambda_S{\tilde S}+\lambda_{\bar S}{\tilde {\bar S}} )  H^+  \nonumber \\
&-\cos\beta \, {\overline{ {\tilde\chi}^-_2} }P_L ( \lambda_S{\tilde S}+\lambda_{\bar S}{\tilde {\bar S}} ) G^- + \sin\beta\, {\overline {{\tilde\chi}^+_2}}P_L ( \lambda_S{\tilde S}+\lambda_{\bar S}{\tilde {\bar S}} )  G^+  +{\rm h.c.}
\end{align}
Then, from Eqs.~(\ref{tildeS}) and (\ref{tildeSbar}),
we get
\begin{align}
-{\cal L}_S=&~{\overline{ {\tilde\chi}^-_2} }(f_{3L} P_L+f_{3R} P_R) {\tilde\chi}^0_3 H^- + {\overline{ {\tilde\chi}^-_2} }(f_{4L} P_L+f_{4R} P_R) {\tilde\chi}^0_4 H^- +{\rm h.c.} \nonumber \\
&+\,(H^-\to G^-,\,\sin\beta\to
  -\cos\beta,\,\cos\beta\to \sin\beta)+{\rm h.c.} +\cdots ,
\end{align}
where
\begin{align}
f_{3L}&=  i \lambda_-\sin\beta, \nonumber\\
f_{3R}&=  - i \lambda_-\cos\beta, \nonumber\\
f_{4L}&= \lambda_+ \sin\beta, \nonumber\\
f_{4R}&=  \lambda_+ \cos\beta.
\end{align}
On the other hand, the interactions between the $W$-boson and singlino-like neutralinos come from the mixing with Higgsinos, given as follows.
\bea
-{\cal L}_V = \frac{g}{\sqrt{2}}\, {\overline{ {\tilde\chi}^-_2} }\gamma^\mu P_L {\tilde H}^0_d W^-_\mu+  \frac{g}{\sqrt{2}}\, {\overline{ {\tilde\chi}^+_2} } \gamma^\mu P_L {\tilde H}^0_u W^+_\mu +{\rm h.c.}
\eea
Then, from Eqs.~(\ref{tildeHd}) and (\ref{tildeHu}), we get the singlino-like interactions to the $W$-boson as
\bea
-{\cal L}_V = {\overline{ {\tilde\chi}^-_2} } \gamma^\mu (g_{3L}
P_L+g_{3R} P_R)  {\tilde \chi}^0_3 W^-_\mu +  {\overline{
    {\tilde\chi}^-_2} } \gamma^\mu (g_{4L} P_L+g_{4R} P_R)  {\tilde
  \chi}^0_4 W^-_\mu +{\rm h.c.}+\cdots ,
\eea
where
\begin{align}
g_{3L}&= \frac{i}{4} g\lambda_- \left(\frac{v_u-v_d}{\mu_{\rm eff}+M_S} -\frac{v_u+v_d}{\mu_{\rm eff}-M_S}\right), \nonumber\\
g_{3R}&= -\frac{i }{4} g\lambda_-\left(\frac{v_u-v_d}{\mu_{\rm eff}+M_S} +\frac{v_u+v_d}{\mu_{\rm eff}-M_S}  \right), \nonumber\\
g_{4L} &=  \frac{ 1}{4} g\lambda_+ \left(\frac{v_u-v_d}{\mu_{\rm eff}-M_S} -\frac{v_u+v_d}{\mu_{\rm eff}+M_S}  \right), \nonumber\\
g_{4R}&= \frac{ 1}{4} g\lambda_+ \left(\frac{v_u-v_d}{\mu_{\rm eff}-M_S} +\frac{v_u+v_d}{\mu_{\rm eff}+M_S}  \right).
\end{align}

Therefore, the magnetic transition dipole moment, generated from charged Higgs, Goldstone, and $W$-boson loops, is given by
\be
{\cal L}_{\rm mdm}=\frac{e f_\chi}{2m_{{\tilde \chi}^0_3}}\,
{\overline {{\tilde \chi}^0_4}}\, i \sigma_{\mu\nu} {\tilde \chi}^0_3
F^{\mu\nu} ,
\ee
where $f_\chi\equiv f^H_\chi +f^G_\chi+  f^W_\chi$ with
\begin{align}
f^H_\chi &=-\frac{\lambda_+\lambda_-}{16\pi^2}\,\cos 2\beta \int^1_0 dx \frac{m^2_{{\tilde\chi}^0_3}\, x(1-x)}{m^2_{{\tilde\chi}^0_3}x^2+(m^2_{H^\pm}-m^2_{{\tilde\chi}^0_3}) x+m^2_{{\tilde\chi}^\pm_2}(1-x)} , \nonumber\\
f^G_\chi&= +\frac{\lambda_+\lambda_-}{16\pi^2}\,\cos 2\beta \int^1_0 dx \frac{m^2_{{\tilde\chi}^0_3}\, x(1-x)}{m^2_{{\tilde\chi}^0_3}x^2+(m^2_W-m^2_{{\tilde\chi}^0_3}) x+m^2_{{\tilde\chi}^\pm_2}(1-x)}, \nonumber\\
f^W_\chi &=  -\frac{\lambda_+\lambda_-}{32\pi^2}\cos 2\beta\frac{m^2_W(\mu^2_{\rm eff}+M^2_S)}{(\mu^2_{\rm eff}-M^2_S)^2} \int^1_0 dx \frac{m^2_{{\tilde\chi}^0_3} \,x(x+2)}{m^2_{{\tilde\chi}^0_3}x^2+(m^2_W-m^2_{{\tilde\chi}^0_3}) x+m^2_{{\tilde\chi}^\pm_2}(1-x)}.
\end{align}
We note that due to the interchange between $\cos\beta$ and $\sin\beta$, the unphysical Goldstone contribution is of the same magnitude but the opposite sign as compared to the charged Higgs contribution.
For $\mu_{\rm eff}\gg M_S$ and $m_{H^\pm}\sim m_{{\tilde\chi}^\pm_2}$, the $W$-boson loops tend to be suppressed by $m_W^2$. But, for $\mu_{\rm eff}\sim M_S$, which is necessary for the co-annihilation of DM as will be discussed in the next section, the $W$-boson loops give a dominant contribution to the magnetic dipole moment of DM.

We take two singlino-like neutralinos to be lighter than Higgsino-like neutralinos and almost degenerate in mass.
Then, the heavier singlino ${\tilde\chi}^0_3$ can decay into the lighter one ${\tilde\chi}^0_4$ through the transition magnetic moment or the mixing with Higgsinos. The decay modes are ${\tilde \chi}^0_3\to {\tilde \chi}^0_4 \gamma$ and ${\tilde\chi}^0_3\to {\tilde\chi}^0_4 \nu {\bar \nu}$, where neutrinos in the latter channel is due to the off-shell $Z$-boson.
The energy of the monochromatic photon coming from ${\tilde \chi}^0_3\to {\tilde \chi}^0_4 \gamma$ is given by $E_\gamma\simeq m_{{\tilde \chi}^0_3}- m_{{\tilde \chi}^0_4}$ for $m_{{\tilde\chi}^0_{3,4}}\gg E_\gamma$.
For $|m_{{\tilde \chi}^0_3}- m_{{\tilde \chi}^0_4}| \ll m_{{\tilde \chi}^0_4}$, the decay rates of the heavier singlino are
\bea
\Gamma({\tilde \chi}^0_3\to {\tilde \chi}^0_4 \gamma)=\frac{e^2 f^2_\chi m_{{\tilde\chi}^0_3}}{\pi} \left(1-\frac{m_{{\tilde\chi}^0_4}}{m_{{\tilde\chi}^0_3}}\right)^3,
\eea
and
\bea
\Gamma({\tilde\chi}^0_3\to {\tilde\chi}^0_4 \,\nu {\bar \nu})= \frac{ |v_{34}|^2 G^2_Fm^5_{{\tilde \chi}^0_3}}{10\pi^3}\, \left(1-\frac{m_{{\tilde \chi}^0_4}}{m_{{\tilde \chi}^0_3}}\right)^5,
\eea
where
\be
v_{34}\equiv N_{31} N_{41}-N_{32} N_{42}\approx -\frac{1}{2} \frac{v^2\cos 2\beta}{M^2_S}\lambda_+\lambda_-.
\ee
Due to an extra factor $(\Delta m_{34})^2$, the decay rate for  ${\tilde\chi}^0_3\to {\tilde\chi}^0_4 \nu {\bar \nu}$ is suppressed as compared to the one for ${\tilde \chi}^0_3\to {\tilde \chi}^0_4 \gamma$.  Thus, it is sufficient to consider only  the decay mode ${\tilde \chi}^0_3\to {\tilde \chi}^0_4 \gamma$ to determine the decay rate of DM.

Suppose that the heavier singlino constitutes a fraction of the total DM by $r\equiv \Omega_{{\tilde \chi}^0_3}/{\Omega_{\rm DM}}$.
Then, for the X-ray line at $3.55\,{\rm keV}$, we need to take the necessary value of the lifetime of the heavier singlino to be $\tau_{{\tilde\chi}^0_3}=0.20$--$1.8\times 10^{28}\,{\rm sec}\,(7.1\,{\rm keV}/m_{{\tilde\chi}^0_3})r$~\cite{xray,mdm}, which is equivalent to $\Gamma_{{\tilde\chi}^0_3}=0.36$ -- $3.3 \times 10^{-52}\,{\rm GeV} \,(m_{{\tilde\chi}^0_3}/7.1\,{\rm keV})r^{-1}$ . For comparably small $\lambda_S$ and $\lambda_{\bar S}$, and a small mass splitting between singlinos, two singlinos contribute to the relic density equally, that is, $r=1/2$.
In Fig.~\ref{allbounds}, we show the parameter space for the mass splitting $\mu_{\rm eff}-M_S$ vs $|\lambda_S|$ or $1-\lambda_{\bar S}/\lambda_S$, satisfying the X-ray line flux (in the region between two black solid lines) and the X-ray line energy (in the blue dashed line).
Therefore, the singlino Yukawa couplings of order $ 10^{-5}$ required for the X-ray line energy is consistent with the X-ray line flux, as far as both Yukawa couplings are of similar size, that is, $\lambda_{\bar S}/\lambda_S\simeq 0.97-0.99$.

\section{Dark matter relic density}
\label{sec:relic}
\noindent
Depending on the singlino Yukawa couplings to Higgsinos, $\lambda_S$ and $\lambda_{\bar S}$, the singlino DM may be in thermal equilibrium with the SM particles due to self-annihilation and/or co-annhiliation with charged and neutral Higgsinos \cite{falkowski}.
The annihilation channels for singlinos are ${\tilde \chi}^0_i {\tilde\chi}^0_j\to f{\bar f}, ZH^0(h^0)$, $W^+ H^-, W^+ W^-$, ${\tilde\chi}^0_i {\tilde \chi}^0_{1,2}\to f {\bar f}$ and ${\tilde\chi}^0_i {\tilde \chi}^\pm_2\to f {\bar f}'$ $(i,j=3,4)$.

In the case with small $\lambda_S$ and $\lambda_{\bar S}$, the self-annihilation cross sections would be too small to make DM in thermal equilibrium with the SM particles.
However, DM can keep in thermal equilibrium until freeze-out, through the scattering off of the SM particles or due to a sizable co-annihilation with neutral or charged Higgsino by crossing symmetry \cite{griest}. In this case, we can obtain a correct relic density for DM, after Higgsinos are decoupled from the SM bath and decay into DM. Therefore, we need $\lambda_S,\,\lambda_{\bar S}\gtrsim 10^{-5}$ for thermal DM \cite{falkowski}.
The mass splitting between singlinos is $3.55\,{\rm keV}$, so it can be ignored in computing the relic density.

The relic abundance is given by
\be
\Omega_{\rm DM}=\frac{8.8\times 10^{-11}\,{\rm GeV}^{-2}}{\sqrt{g_*}\int^\infty_{x_f} dx \langle \sigma_{\rm eff} v\rangle x^{-2}} , \label{relic}
\ee
where $g_*$ is the effective number of relativistic degrees of freedom at freeze-out and $x\equiv m_{\rm DM}/T$ which read $x_f\approx 20$ at freeze-out temperature.
The effective cross section is a weighted average of the annihilation cross sections  for the co-annihilating particles and is given \cite{griest,falkowski} by
\be
\langle \sigma_{\rm eff} v \rangle = \frac{\sum_{i,j} \sigma_{ij} w_i
  w_j}{(\sum_i w_i)^2 } ,
\ee
where
\be
w_i\equiv \left(1+\Delta_i\right)^{3/2} \, e^{-x \Delta_i}, \quad \Delta_i\equiv \frac{m_i-m_{\rm DM}}{m_{\rm DM}},
\ee
and $\langle \sigma_{ij} v\rangle=\sigma_{ij} x^{-n}$ with $n=0$ (1) for $s$-wave ($p$-wave) annihilation.

\begin{figure}
  \begin{center}
    \includegraphics[height=0.48\textwidth]{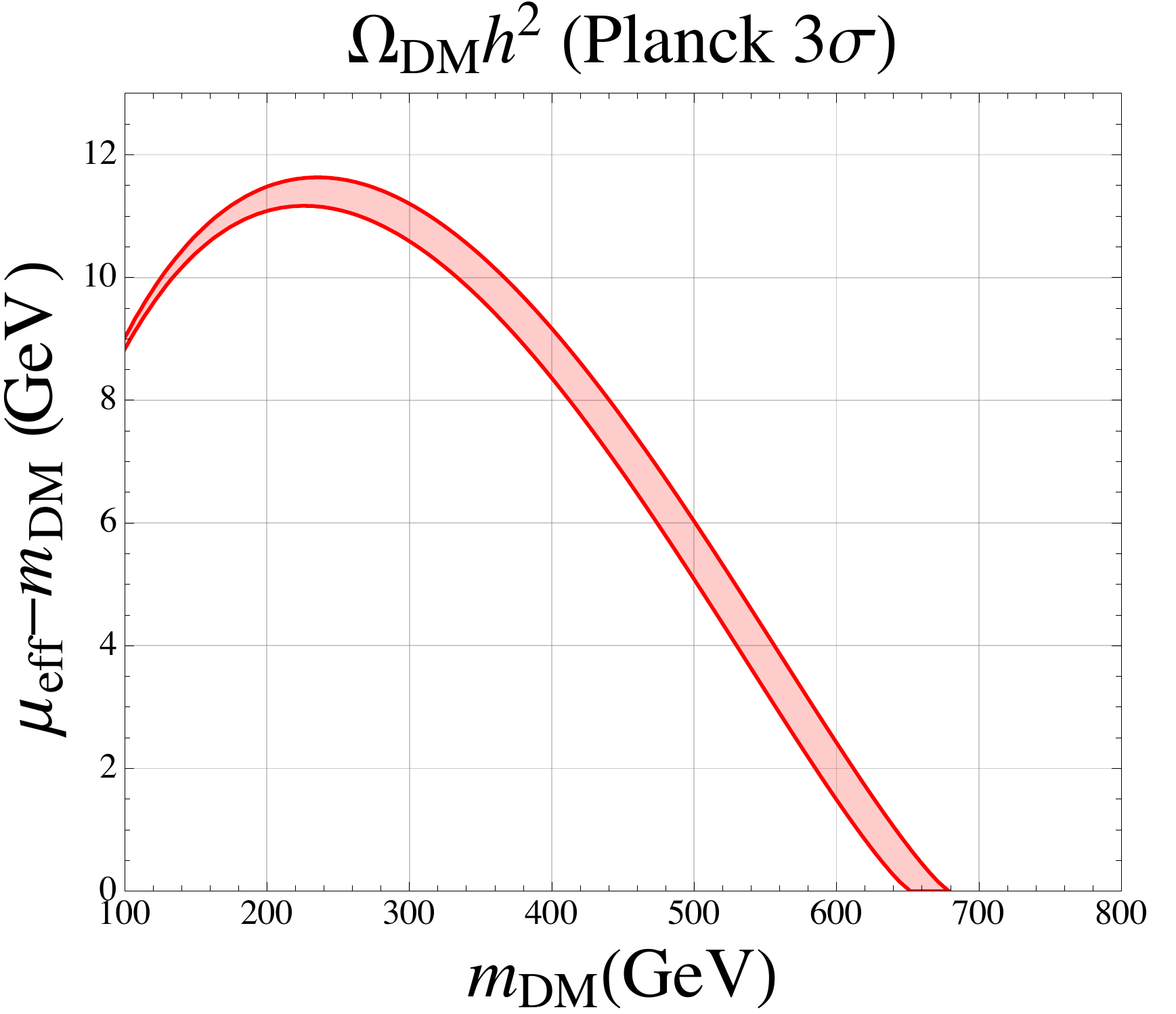}
  \end{center}
  \caption{ Parameter space of  $m_{\rm DM}$ vs $\mu_{\rm eff}-M_S$, satisfying the relic density within Planck $3\sigma$~\cite{planck}.
  }
  \label{relicdm}
\end{figure}

For small $\lambda_S$ and $\lambda_{\bar S}$, the effective annihilation cross section is dominated by the annihilation of neutral or charged Higgsino so we find that
\be
\langle \sigma_{\rm eff} v\rangle =\frac{1}{4(\sum_i w_i)^2}
(\sigma_{{\tilde\chi}^0_1{\tilde\chi}^0_1} w^2_{{\tilde\chi}^0_1}
+\sigma_{{\tilde\chi}^0_2{\tilde\chi}^0_2}
w^2_{{\tilde\chi}^0_2}+2\sigma_{{\tilde\chi}^\pm_1{\tilde\chi}^\mp_1}
w^2_{{\tilde\chi}^\pm_1}
+2\sigma_{{\tilde\chi}^0_1{\tilde\chi}^\pm_1}w_{{\tilde\chi}^0_1}
w_{{\tilde\chi}^\pm_1}+2\sigma_{{\tilde\chi}^0_2{\tilde\chi}^\pm_1}w_{{\tilde\chi}^0_2}
w_{{\tilde\chi}^\pm_1}  ) ,
\ee
where
\be
\sum_i w_i=\frac{1}{2}w_{{\tilde\chi}^0_1}+\frac{1}{2}w_{{\tilde\chi}^0_2}+ w_{{\tilde\chi}^\pm_1}  +1.
\ee
In the limit of decoupled gauginos, $w\equiv w_{{\tilde\chi}^0_1}\approx w_{{\tilde\chi}^0_2}\approx w_{{\tilde\chi}^\pm_1} $, so we obtain \cite{falkowski}
\be
\langle \sigma_{\rm eff} v\rangle =\frac{\sigma_{{\tilde H}{\tilde H}}w^2 }{(w+\frac{1}{2})^2} ,\label{xection}
\ee
where
\be
\sigma_{{\tilde H}{\tilde H}}= \frac{81 g^4+12 g^{\prime 2} g^2+43 g^{\prime 4}}{2048\pi \mu^2_{\rm eff}},\quad  w =\left(\frac{\mu_{\rm eff}}{M_S}\right)^{3/2} {\rm exp}\left[-x \Big(\frac{\mu_{\rm eff}}{M_S}-1\Big)\right].
\ee
In this case, the relic density can be determined from Eqs.~(\ref{relic}) and (\ref{xection}) and the parameter space for the DM mass and the mass splitting between Higgsinos and DM is shown in Fig.~\ref{relicdm}. Below the red region, the relic density is smaller than the lower end of the Planck $3\sigma$ values so the parameter space in that region is consistent with those obtained for explaining the X-ray line in Fig.~\ref{allbounds}.

If the Higgsino mass splitting is not ignorable, the contribution of the lighter neutral/charged Higgsino to the effective annihilation cross section gets larger, while the one of the heavier Higgsino gets smaller. But, overall, the effective annihilation cross section would increase due to the lighter charged Higgsino. Therefore, the difference between the averaged Higgsino mass and the singlino can be larger than $\mu_{\rm eff}-M_S$ as shown in Fig.~\ref{relicdm}.

\section{Collider searches}
\noindent
When the gauginos are heavy enough, the mass splitting between neutral Higgsinos is of about ${\rm keV}$ scale, being as small as the one between singlinos, and the charged Higgsino is almost degenerate in mass with the neutral Higgsinos.  On the other hand, the difference between Higgsino and singlino masses should be less than about $10\,{\rm GeV}$ for the co-annihilation with singlino DM.
Thus, neutral Higgsinos and charged Higgsino decay dominantly into singlinos with three-body modes such as ${\tilde\chi}^0_2\to {\tilde\chi}^0_1 \nu {\bar \nu}$, ${\tilde\chi}^0_{1,2}\to {\tilde\chi}^0_{3,4} Z^\ast$ and ${\tilde\chi}^\pm_2\to {\tilde\chi}^0_{3,4} W^{\pm \ast} $, as well as the modes containing Higgs fields.
Since the singlino Yukawa couplings are small for the X-ray line, the charged/neutral Higgsinos decay modes into singlino plus off-shell $W/Z$ leave displaced vertex plus missing energy~\cite{falkowski}.
When the decay length of the charged Higgsino is between about $1\,{\rm cm}$ and $100\,{\rm m}$,  there is a bound on the mass of the charged Higgsino from the disappearing tracks at the LHC \cite{charge track}.  In our case, when the charged Higgsino decays mostly into singlino plus off-shell $W$, the decay length of the charged Higgsino is about {\cal O}({\rm mm}-{\rm m}) as in Ref.~\cite{falkowski}. Thus, a certain parameter space of a small mass splitting can be constrained. 
However, the LEP, Tevatron and the LHC Run I~\cite{lhc-displaced} are not sensitive enough to rule out the neutral Higgsinos. 

When Higgsinos have a sizable mass splitting due to the non-decoupling effect of gauginos, the heavier neutral Higgsino can decay into the lighter neutral or charged Higgsino with a sizable branching fraction, and the charged Higgsino can decay into neutral Higgsinos as well. In this case, since Higgsinos have gauge interactions, there is no displaced vertex. However, depending on the mass splitting of the Higgsinos, missing energy plus collimated leptons at the primary vertices can be a signature.
In order for the gauginos not to give a large contribution to the singlino mass splitting, their contribution to the Higgsino mass splitting is less than $5\,{\rm GeV}$. In this case, the situation would be better due to larger efficiency of the lepton momentum cuts, as compared to the Higgsinos with ${\rm keV}$ mass splitting.  The detailed discussion on the search for almost degenerate Higgsinos is outside the scope of this work, so it is left for a future publication~\cite{work2}.

\section{Conclusions}
\noindent
We have considered a Dirac singlet fermion or singlinos with small mass splitting as thermal DM in the SUSY Higgs-portal model. In order to explain the X-ray line excess observed from the sky, we introduced small singlino Yukawa couplings with Higgses and Higgsinos, the SUSY version of Higgs-portal couplings, and showed that the mass splitting of $3.55\,{\rm keV}$ is made and at the same time a tiny magnetic transition dipole moment between the Majorana components of the singlino is generated.
The singlino mass splitting requires gaugino masses to be heavier than about $1\,{\rm TeV}$, leading to almost degenerate Higgsinos. The thermal production of  the singlino DM restricts the Higgsino masses to be not greater than about $10\,{\rm GeV}$ as compared to the singlino masses.
New search strategies for almost degenerate Higgsinos at the LHC Run II and future colliders are needed to probe the SUSY Higgs-portal models.

\section*{Acknowledgments}

The work of HML is supported in part by Basic Science Research Program through the National Research Foundation of Korea (NRF) funded by the Ministry of Education, Science and Technology (2013R1A1A2007919). MP was supported by World Premier International Research Center Initiative (WPI Initiative), MEXT, Japan and acknowledges support from the Korea Ministry of Science, ICT and Future Planning, Gyeongsangbuk-Do and Pohang City for Independent Junior Research Groups at the Asia Pacific Center for Theoretical Physics.

%


\begin{thebibliography}{999}



\bibitem{xray}
  E.~Bulbul, M.~Markevitch, A.~Foster, R.~K.~Smith, M.~Loewenstein and S.~W.~Randall,
  Astrophys.\ J.\  {\bf 789}, 13 (2014)
  [arXiv:1402.2301 [astro-ph.CO]];
\bibitem{Boyarsky:2014jta}
  A.~Boyarsky, O.~Ruchayskiy, D.~Iakubovskyi and J.~Franse,
  Phys.\ Rev.\ Lett.\  {\bf 113}, no. 25, 251301 (2014)
  [arXiv:1402.4119 [astro-ph.CO]];
  A.~Boyarsky, J.~Franse, D.~Iakubovskyi and O.~Ruchayskiy,
  arXiv:1408.2503 [astro-ph.CO].


\bibitem{debate}
  T.~E.~Jeltema and S.~Profumo,
  arXiv:1408.1699 [astro-ph.HE];
  A.~Boyarsky, J.~Franse, D.~Iakubovskyi and O.~Ruchayskiy,
  arXiv:1408.4388 [astro-ph.CO];
  E.~Bulbul, M.~Markevitch, A.~R.~Foster, R.~K.~Smith, M.~Loewenstein and S.~W.~Randall,
  arXiv:1409.4143 [astro-ph.HE];
  T.~Jeltema and S.~Profumo,
  arXiv:1411.1759 [astro-ph.HE];
  E.~Carlson, T.~Jeltema and S.~Profumo,
  arXiv:1411.1758 [astro-ph.HE].


\bibitem{dmmodels}
  H.~Ishida, K.~S.~Jeong and F.~Takahashi,
  Phys.\ Lett.\ B {\bf 732}, 196 (2014)
  [arXiv:1402.5837 [hep-ph]];
  D.~P.~Finkbeiner and N.~Weiner,
  arXiv:1402.6671 [hep-ph];
  T.~Higaki, K.~S.~Jeong and F.~Takahashi,
  Phys.\ Lett.\ B {\bf 733}, 25 (2014)
  [arXiv:1402.6965 [hep-ph]];
  J.~Jaeckel, J.~Redondo and A.~Ringwald,
  Phys.\ Rev.\ D {\bf 89}, no. 10, 103511 (2014)
  [arXiv:1402.7335 [hep-ph]];
  H.~M.~Lee, S.~C.~Park and W.~I.~Park,
  Eur.\ Phys.\ J.\ C {\bf 74}, no. 9, 3062 (2014)
  [arXiv:1403.0865 [astro-ph.CO]];
  R.~Krall, M.~Reece and T.~Roxlo,
  JCAP {\bf 1409}, 007 (2014)
  [arXiv:1403.1240 [hep-ph]];
  J.~C.~Park, S.~C.~Park and K.~Kong,
  Phys.\ Lett.\ B {\bf 733}, 217 (2014)
  [arXiv:1403.1536 [hep-ph]];
  M.~T.~Frandsen, F.~Sannino, I.~M.~Shoemaker and O.~Svendsen,
  JCAP {\bf 1405}, 033 (2014)
  [arXiv:1403.1570 [hep-ph]];
  K.~Y.~Choi and O.~Seto,
  Phys.\ Lett.\ B {\bf 735}, 92 (2014)
  [arXiv:1403.1782 [hep-ph]];
  S.~Baek and H.~Okada,
  arXiv:1403.1710 [hep-ph];
  M.~Cicoli, J.~P.~Conlon, M.~C.~D.~Marsh and M.~Rummel,
  Phys.\ Rev.\ D {\bf 90}, no. 2, 023540 (2014)
  [arXiv:1403.2370 [hep-ph]];
  F.~Bezrukov and D.~Gorbunov,
  Phys.\ Lett.\ B {\bf 736}, 494 (2014)
  [arXiv:1403.4638 [hep-ph]];
  C.~Kolda and J.~Unwin,
  Phys.\ Rev.\ D {\bf 90}, no. 2, 023535 (2014)
  [arXiv:1403.5580 [hep-ph]];
  R.~Allahverdi, B.~Dutta and Y.~Gao,
  Phys.\ Rev.\ D {\bf 89}, no. 12, 127305 (2014)
  [arXiv:1403.5717 [hep-ph]];
  N.-E.~Bomark and L.~Roszkowski,
  Phys.\ Rev.\ D {\bf 90}, no. 1, 011701 (2014)
  [arXiv:1403.6503 [hep-ph]];
  S.~P.~Liew,
  JCAP {\bf 1405}, 044 (2014)
  [arXiv:1403.6621 [hep-ph]];
  Z.~Kang, P.~Ko, T.~Li and Y.~Liu,
  arXiv:1403.7742 [hep-ph];
  S.~V.~Demidov and D.~S.~Gorbunov,
  Phys.\ Rev.\ D {\bf 90}, no. 3, 035014 (2014)
  [arXiv:1404.1339 [hep-ph]];
  F.~S.~Queiroz and K.~Sinha,
  Phys.\ Lett.\ B {\bf 735}, 69 (2014)
  [arXiv:1404.1400 [hep-ph]];
  E.~Dudas, L.~Heurtier and Y.~Mambrini,
  Phys.\ Rev.\ D {\bf 90}, no. 3, 035002 (2014)
  [arXiv:1404.1927 [hep-ph]];
  K.~S.~Babu and R.~N.~Mohapatra,
  Phys.\ Rev.\ D {\bf 89}, no. 11, 115011 (2014)
  [arXiv:1404.2220 [hep-ph]];
  J.~M.~Cline, Y.~Farzan, Z.~Liu, G.~D.~Moore and W.~Xue,
  Phys.\ Rev.\ D {\bf 89}, no. 12, 121302 (2014)
  [arXiv:1404.3729 [hep-ph]];
  S.~Chakraborty, D.~K.~Ghosh and S.~Roy,
  JHEP {\bf 1410} (2014) 146
  [arXiv:1405.6967 [hep-ph]];
  K.~Cheung, W.~C.~Huang and Y.~L.~S.~Tsai,
  arXiv:1411.2619 [hep-ph].



\bibitem{mdm}
  H.~M.~Lee,
  Phys.\ Lett.\ B {\bf 738} (2014) 118
  [arXiv:1404.5446 [hep-ph]].




\bibitem{mdmothers}
  K.~P.~Modak,
  arXiv:1404.3676 [hep-ph];
  C.~Q.~Geng, D.~Huang and L.~H.~Tsai,
  JHEP {\bf 1408} (2014) 086
  [arXiv:1406.6481 [hep-ph]];
  C.~W.~Chiang and T.~Yamada,
  JHEP {\bf 1409} (2014) 006
  [arXiv:1407.0460 [hep-ph]];
  J.~M.~Cline and A.~R.~Frey,
  JCAP {\bf 1410} (2014) 10,  013
  [arXiv:1408.0233 [hep-ph]];
  S.~Baek,
  arXiv:1410.1992 [hep-ph];
  S.~Patra, N.~Sahoo and N.~Sahu,
  arXiv:1412.4253 [hep-ph];
  G.~Arcadi, L.~Covi and F.~Dradi;
  arXiv:1412.6351 [hep-ph].
  A.~Berlin, A.~DiFranzo and D.~Hooper,
  arXiv:1501.03496 [hep-ph].



\bibitem{falkowski}
  A.~Falkowski, Y.~Hochberg and J.~T.~Ruderman,
  JHEP {\bf 1411}, 140 (2014)
  [arXiv:1409.2872 [hep-ph]].



\bibitem{dnmssm}
  X.~Lu, H.~Murayama, J.~T.~Ruderman and K.~Tobioka,
  Phys.\ Rev.\ Lett.\  {\bf 112} (2014) 191803
  [arXiv:1308.0792 [hep-ph]].


\bibitem{kai}
  A.~Kaminska, G.~G.~Ross, K.~Schmidt-Hoberg and F.~Staub,
  JHEP {\bf 1406} (2014) 153
  [arXiv:1401.1816 [hep-ph]].




\bibitem{Nmdm}
  D.~Schmidt, T.~Schwetz and T.~Toma,
  Phys.\ Rev.\ D {\bf 85} (2012) 073009
  [arXiv:1201.0906 [hep-ph]].


\bibitem{jekim}
  W.~S.~Cho, J.~H.~Huh, I.~W.~Kim, J.~E.~Kim and B.~Kyae,
  Phys.\ Lett.\ B {\bf 687} (2010) 6
   [Erratum-ibid.\ B {\bf 694} (2011) 496]
  [arXiv:1001.0579 [hep-ph]].


\bibitem{haber}
  H.~E.~Haber and D.~Wyler,
  Nucl.\ Phys.\ B {\bf 323} (1989) 267.


\bibitem{griest}
  K.~Griest and D.~Seckel,
  Phys.\ Rev.\ D {\bf 43} (1991) 3191.

\bibitem{work2}
H.~M.~Lee, C.~B.~Park and M.~Park, To appear.

\bibitem{planck}
  P.~A.~R.~Ade {\it et al.}  [Planck Collaboration],
  Astron.\ Astrophys.\  {\bf 571} (2014) A16
  [arXiv:1303.5076 [astro-ph.CO]].


\bibitem{chargetrack}
  [CMS Collaboration],
  JHEP {\bf 01} (2015) 096
  [arXiv:1411.6006 [hep-ex]].



\bibitem{lhc-displaced}
  G.~Aad {\it et al.}  [ATLAS Collaboration],
  Phys.\ Rev.\ Lett.\  {\bf 108} (2012) 251801
  [arXiv:1203.1303 [hep-ex]];
  G.~Aad {\it et al.}  [ATLAS Collaboration],
  Phys.\ Lett.\ B {\bf 721} (2013) 32
  [arXiv:1210.0435 [hep-ex]];
  ATLAS collaboration,
  ATLAS-CONF-2013-092, ATLAS-COM-CONF-2013-108;
  G.~Aad {\it et al.}  [ATLAS Collaboration],
  JHEP {\bf 1411} (2014) 088
  [arXiv:1409.0746 [hep-ex]];
  CMS Collaboration,
  CMS-PAS-EXO-12-037;
  CMS Collaboration,
  CMS-PAS-EXO-12-038;
  CMS Collaboration,
  CMS-PAS-B2G-12-024.


\end{thebibliography}
\end{document}